\documentclass[aps,twocolumn,preprintnumbers,amsmath,nofootinbib,amssymb]{revtex4}

\usepackage{graphics}
\usepackage{epsfig,epsf,psfrag}
\usepackage{times}
\usepackage{float}
\usepackage{color}
\usepackage{amsfonts,amssymb,stmaryrd,latexsym,amsmath}
\usepackage{multirow}
\usepackage{array}
\usepackage{hhline}
\usepackage{booktabs}
\usepackage{enumerate}
\usepackage{slashed}
\usepackage{subfigure}
\usepackage[colorlinks, citecolor=blue,anchorcolor=red,menucolor=red, linkcolor=red,filecolor=red,runcolor=red,urlcolor=blue,frenchlinks=red]{hyperref}
\usepackage{epstopdf}

\begin{document}

\bibliographystyle{unsrt}

\title{Triangle singularity in $B^0\to \pi^- K^+ X(3872)$ via the $D_{s1}\bar{D} D^*$ loop and possible precise measurement of the $X(3872)$ mass}

\author{Mao-Jun~Yan$^{1}$}
\author{Ying-Hui~Ge$^{2}$}
\author{Xiao-Hai~Liu$^{2}$}~\email{xiaohai.liu@tju.edu.cn}

\affiliation{$^{1}$CAS Key Laboratory of Theoretical Physics, Institute of Theoretical Physics, \\Chinese Academy of Sciences, Beijing 100190, China\\
$^{2}$Center for Joint Quantum Studies and Department of Physics, School of Science, Tianjin University, Tianjin 300350, China}

\date{\today}

\begin{abstract}

 We investigate the $B^0\to \pi^- K^+ X(3872)$ decay via the $D_{s1}(2536)\bar{D} D^*$ rescattering diagram. The line shape of the $K^+X(3872)$ distribution curve around $D_{s1}(2536)\bar{D}$ threshold is very sensitive to the $X(3872)$ mass because the triangle singularity (TS) can be generated from the loop. By means of this characteristic, we can determine whether the $X(3872)$ mass is below or above the $D^{\ast 0}\bar{D}^0$ threshold with high precision. The narrowness of $D_{s1}(2536)$ in the loop is one of the key reasons why the TS mechanism of measuring the $X(3872)$ mass may work. The $X(3872)$ width impact on the $K^+X(3872)$ line shape is also crucial in the TS mechanism. If the width is as large as 1 MeV, the proposed method of measuring the $X(3872)$ mass would be ruined.

\end{abstract}
\maketitle

\section{Introduction}

The threshold cusp and triangle singularity (TS) have been known for many years. They are kinematic singularities of the $S$-matrix and their locations are determined by kinematic variables instead of the interaction strength, which are different from the pole singularities corresponding to hadrons whose origin is dynamical. The square-root branch point of the amplitude at the normal two-body threshold can produce cusp in the energy distribution. The more complicated TS is a logarithmic Landau singularity of the amplitude, which can appear in the physical region due to three on-shell intermediate particles in the loop diagram. Observable effects produced from the threshold cusp and TS, especially the latter, have received more and more attention in recent years. Although some observable effects induced by the TS have been noticed as early as the 1960s, there were limited processes that were accessible in experiments at that time. With the development of experiments, there have been quite a few exotic phenomena that are suggested to result from the TS. We refer to Ref.~\cite{Guo:2019twa} for a recent review about the threshold cusp and TS in hadronic reactions.

One of the significant high-energy experimental achievements in recent years is the discovery of dozens of exotic hadrons, many of which are also named as $XYZ$ particles (see Refs.~\cite{PDG:2022,Guo:2017jvc,Kalashnikova:2018vkv,Brambilla:2019esw,Chen:2016qju,Esposito:2016noz,Olsen:2017bmm,Lebed:2016hpi,Ali:2017jda,Liu:2019zoy} for a review). An intriguing feature of these exotic states is that many of them are located close to two-hadron thresholds. This is the reason why many of them are regarded as hadronic molecules in numerous papers. Among those candidates of hadronic molecules, the $X(3872)$ (aka $\chi_{c1}(3872)$ in Ref.~\cite{PDG:2022}) could be the most famous one. It is the first unconventional charmonium-like state observed in the experiment~\cite{Abe:2003hq}. Its $J^{PC}$ quantum numbers are determined to be $1^{++}$ which thus could be the candidate for the quark model state $\chi_{c1}(2P)$. Its preferred decay mode of $\gamma\psi(2S)$ over $\gamma J/\psi$ also favors the $\chi_{c1}(2P)$ assignment. Furthermore, its large production rate at LHC~\cite{ATLAS:2016kwu,CMS:2013fpt,LHCb:2011zzp} and Tevatron~\cite{D0:2004zmu} implies that it may contain a compact component. However, its mass is just in the vicinity of the ${D}^{*0}\bar{D}^0 $ ($D^0 \bar{D}^{*0}$) threshold, which is far from the quark model prediction. The 2022 Particle Data Group (PDG) world-average value is $m_X=3871.65\pm 0.06$ MeV~\cite{PDG:2022}. The ${D}^{*0}\bar{D}^0 $ ($D^0 \bar{D}^{*0}$) threshold is $m_{D^0}+m_{D^{*0}}=3871.69\pm 0.07$ MeV. Then the difference is 
\begin{equation}
	\delta_X\equiv m_{D^0} + m_{D^{\ast 0}}-m_X=0.04\pm 0.09\ \mbox{MeV}.
\end{equation}
The incredible closeness of the $m_X$ to the ${D}^{*0}\bar{D}^0 $ ($D^0 \bar{D}^{*0}$) threshold together with the large branching ratio of the $X(3872)$ into ${D}^{*0}\bar{D}^0 +c.c.$ suggest that the natural explanation of $X(3872)$ could be a hadronic molecule. In this case, the $\delta_X$ can be understood as the binding energy. One can find experimental evidences for both the compact state and hadronic molecule interpretations of the $X(3872)$. Although the $X(3872)$ has been well experimentally established by now, its intrinsic structure is still quite puzzling. 

From the above $\delta_X$ value one can see the $m_X$ is still indistinguishable from the ${D}^{*0}\bar{D}^0 $ ($D^0 \bar{D}^{*0}$) threshold at current levels of precision, i.e., whether the $X(3872)$ mass is above or below ${D}^{*0}\bar{D}^0 $ ($D^0 \bar{D}^{*0}$) threshold is still unknown, while a high accuracy mass determination of the $X(3872)$ is very important in understanding its underlying structure. In a recent paper Ref.~\cite{Guo:2019qcn} a novel method was proposed to measure the $X(3872)$ mass precisely by measuring the $\gamma X(3872)$ line shape. In the rescattering process $D^{\ast 0}\bar{D}^{\ast 0}\to \gamma X(3872)$, where $D^{\ast 0}\bar{D}^{\ast 0}$ would be produced by a short-distance source, the line shape of $\gamma X(3872)$ invariant mass spectrum is very sensitive to the $m_X$ or the binding energy $\delta_X$ defined above. This is because the TS location of the rescattering diagram is rather sensitive to the particle masses involved. For $\delta_X > 0$ and $\delta_X \le 0$, the corresponding line shapes show a significant discrepancy. 

In Refs.~\cite{Guo:2019qcn,Sakai:2020crh}, the $D^{*0} \bar{D}^{*0}$ pair produced from the short-distance source is set to be in the $S$-wave. In Refs. \cite{Braaten:2019gfj,Braaten:2019gwc}, the authors implement this TS mechanism and give a possible reaction $e^+ e^- \to \gamma X(3872)$ via the $D^{*0} \bar{D}^{*0}$ rescattering in the $P$-wave. Although the $P$-wave scattering may smooth the TS peak to some extent, this kind of measurement may be available at current electron-positron colliders. Another similar method by measuring the $\pi X(3872)$ invariant mass spectrum in the $B\to \pi K X(3872)$ process is also suggested in Refs.~\cite{Braaten:2019yua,Sakai:2020ucu}. Besides, the production of the double-charm tetraquark candidate $T_{cc}^+(3875)$ via a similar $D^*D^*D$ triangle rescattering diagram was also studied in Ref.~\cite{Braaten:2022elw}.

The TSs in the above studies concerning the $X(3872)$ production are all developed from the $D^*\bar{D}^*D$ triangle loops, where $D^*\bar{D}^*$ scatter into $\gamma X(3872)$ or $\pi X(3872)$ via exchanging the $D$ meson.  One important reason why this novel method of measuring the $X(3872)$ mass may work is that the $D^*$ ($\bar{D}^*$) meson in the triangle diagram is quite narrow, which leads to that the line shape of $\gamma X(3872)$ or $\pi X(3872)$ spectrum is sensitive to the mass of $X(3872)$. Besides the $D^*\bar{D}^*D$ triangle loops, a similar scenario may also appear in other processes. In this work, we suggest to measure the $KX(3872)$ distribution in $B\to \pi KX(3872)$ via the $D_{s1}\bar{D}D^*$ loop, which possesses some special advantages for the determination of $m_X$.

\section{The Model}\label{SecII}
\subsection{TS mechanism}

\begin{figure}[htb]
	\centering
	\includegraphics[width=0.43\textwidth]{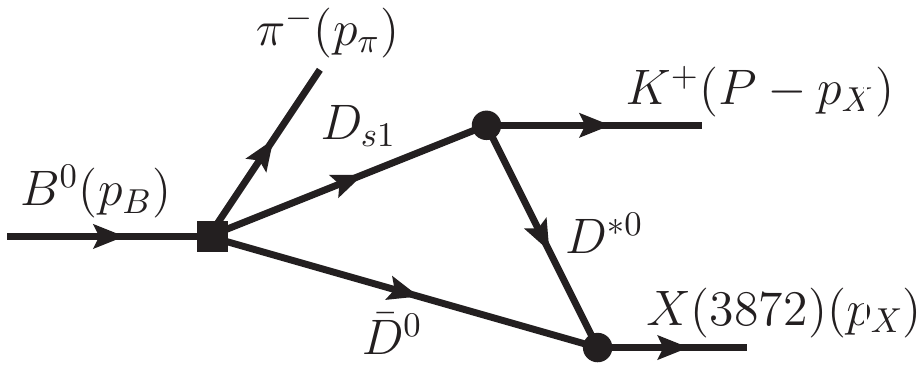}
	\caption{$B^0 \to \pi^- K^+ X(3872)$ via the $D_{s1}(2536)^+\bar{D}^0 D^{\ast 0}$ triangle rescattering diagram. We define the the invariant $M_{KX}^2\equiv P^2\equiv (p_B-p_\pi)^2$.}\label{triangleDiagram}
\end{figure}

The $B^0\to \pi^- K^+ X(3872)$ is one of the reactions where the $X(3872)$ is discovered, of which the branching fraction is around $(2.1\pm 0.8)\times 10^{-4}$~\cite{ParticleDataGroup:2020ssz,Belle:2015qeg}. We notice that this process may receive contributions from the triangle diagram displayed in Fig.~\ref{triangleDiagram}. In this rescattering process, the $B^0\to \pi^- D_{s1}(2536) \bar{D}^0$ is a Cabibbo-favored decay, and the $D_{s1}(2536)$ mainly decays into $D^* K$. Therefore we can expect this rescattering may play a role in $B^0\to \pi^- K^+ X(3872)$. Furthermore, the intriguing feature of this rescattering process is that the three intermediate particles can be (nearly) on-shell simultaneously in some kinematic regions, and a TS located close to the physical boundary in the complex energy plane of the amplitude can develop from this $D_{s1}\bar{D}D^*$ loop. As a result the transition amplitude of $B^0\to \pi^- K^+ X(3872)$ will be enhanced in some areas and a TS peak can be expected to arise in the $K^+ X(3872)$ invariant mass spectrum.

Assuming the $X(3872)$ mass $m_X$ is not fixed, the location of the TS in the $m_X$ or $M_{KX}$ complex plane can be determined by solving the Landau equation~\cite{Coleman:1965xm,Landau:1959fi}. In terms of Eqs.~(3) and (4) of Ref.~\cite{Liu:2015taa} derived from the Landau equation and a dispersion analysis, we can obtain the TS window corresponding to Fig.~\ref{triangleDiagram}:
\begin{eqnarray}
	m_X&\in& \left[3871.69,~3875.24  \right]~\rm{MeV},\label{eq:2}\\
	M_{KX} & \in & \left[4399.93,~4403.66\right]~\rm{MeV},\label{eq:3}
\end{eqnarray}
where the central mass values of relevant mesons from Ref.~\cite{PDG:2022} are adopted. The meaning of the above window is : When $m_X$ increases from $3871.69$ MeV, i.e., the ${D}^{*0}\bar{D}^0 $ threshold, to $3875.24$ MeV, the TS in $M_{KX}$ moves from $4403.66$ to $4399.93$ MeV; Vice versa, when $M_{KX}$ increase from $4399.93$ MeV, i.e., the $D_{s1}(2536)\bar{D}^0$ threshold, to $4403.66$ MeV, the TS in $m_X$ moves from $3875.24$ to $3871.69$ MeV.  We also refer to Refs.~\cite{Landau:1959fi,Coleman:1965xm,Bronzan:1963mby,Aitchison:1964zz,Liu:2015taa} for more detailed discussion on the locations of the TS in various kinematic configurations.

For the $D^*\bar{D}^*D$ loop mentioned in the introduction, ignoring the $D^*$ width, when $\delta_X=0$, the TS position of $M_{\gamma X}$ is about 2.7 MeV larger than the $D^{*}\bar{D}^{*}$ threshold, while that of $M_{\pi X}$ is 0.3 MeV. For the $D_{s1}\bar{D}D^*$ loop, ignoring the $D_{s1}(2536)$ width, when $\delta_X=0$, the TS position of $M_{KX}$ is 3.7 MeV larger than the $D_{s1}\bar{D}^0$ threshold. Larger gap between the TS position and pertinent threshold indicates that the line shape could be more sensitive to the $X(3872)$ mass compared with the $D^*\bar{D}^*D$ loop. Besides, the charm-strange meson $D_{s1}(2536)$ is also very narrow. The PDG average value is $\Gamma(D_{s1}(2536)^\pm)=0.92\pm 0.05$ MeV~\cite{PDG:2022}. The narrowness of intermediate particles in the triangle diagram is one of the key reasons why the TS mechanism of measuring the $X(3872)$ mass may work. 


\subsection{Amplitude of $B^0\to \pi^- K^+ X(3872)$}

When employing the TS mechanism to determine the $X(3872)$ mass, we are interested in the line shape of $M_{KX}$ distributions in the vicinity of $D_{s1}(2536)\bar{D}^0$ threshold, therefore we only take into account the amplitude involving the lowest angular momentum between $D_{s1}(2536)$ and $\bar{D}^0$. The non-relativistic amplitude of $B^0\to \pi^- D_{s1}\bar{D}^0$ reads
\begin{eqnarray}\label{vertex1}
	t_{B^0\to \pi^- D_{s1}\bar{D}^0}&=&\mathcal{C}_1\vec{\epsilon}^* (D_{s1})\cdot\vec{p}_{\pi} ,
\end{eqnarray}
 with $\vec{P}=-\vec{p}_{\pi}+\vec{p}_B=0$ in the $D_{s1}(2536)\bar{D}^0$ c.m. frame. The coupling constant $\mathcal{C}_1$ can be determined from the experimental data. However, the branching fraction of $B^0\to \pi^- D_{s1}(2536)\bar{D}^0$ is not known yet. The experiment gives $\mathcal{B}(B^0\to D_{s1}(2536)^+ D^{*-})\times \mathcal{B}(D_{s1}(2536)^+\to (D^{*0} K^+ +D^{*+}K^0))=(5.0\pm 1.4)\times 10^{-4}$~\cite{PDG:2022}. Assuming the $\pi^- \bar{D}^0$ states in $B^0\to \pi^- D_{s1}(2536)\bar{D}^0$ are fully from the $D^{*-}\to \pi^- \bar{D}^0$ decays, we estimate the coupling constant $\mathcal{C}_1 \approx 7.0\times 10^{-6}$ using the experimental central values.

The $D_{s1}(2536)$ mainly decays into $D^{\ast}K$ in relative S-wave, and the amplitude reads
\begin{eqnarray}\label{Ds1DstarK}
	t_{D_{s1}\to D^{\ast 0}K^+}&=&g_{D_{s1}D^*K}\vec{\epsilon}(D_{s1})\cdot\vec{\epsilon}^*(D^{\ast 0}),
\end{eqnarray}
where the coupling $g_{D_{s1}D^*K}$ can be determined from the experimental data. Using the central values of the particle masses and branching fraction from Ref.~\cite{PDG:2022}, we obtain $g_{D_{s1}D^*K}\approx 0.78\ \mbox{GeV}$.
Actually, this $S$-wave decay mode is supposed to be suppressed by the heavy quark spin symmetry (HQSS). On the other hand, the HQSS is preserved for the $D^*K$ $D$-wave decay mode, but this mode is highly suppressed by the limited phase space, since the $D^*K$ threshold is rather close to the $D_{s1}(2536)$ mass. These reasons lead to that the $D_{s1}(2536)$ is so narrow, and we can take advantage of this characteristic to make the TS mechanism work. For another charmed-strange meson $D_{s2}(2573)$, although it can also decay into $D^*K$ and its mass is just a little larger than that of $D_{s1}(2536)$, its width is about 16.9 MeV, which is much larger than $\Gamma_{D_{s1}(2536)}$. We therefore do not take into account the contribution from the $D_{s2}\bar{D}D^*$  loop in this work.

For the fusion of $\bar{D}^0$ and $D^{*0}$ into the $X(3872)$, the amplitude can be written as
\begin{eqnarray}\label{XDDstar}
t_{XD^0 \bar{D}^{\ast 0}}&=& \frac{g_X}{2}\vec{\epsilon}(\bar{D}^{\ast 0})\cdot\vec{\epsilon}^*(X).
\end{eqnarray}
Supposing the $X(3872)$ is a pure hadronic molecule, the coupling $g_X$ can be estimated by using the Weinberg compositeness condition~\cite{Weinberg:1965zz,Baru:2003qq,Gamermann:2009uq}, which gives
\begin{eqnarray}\label{gX}
g_X^2=\frac{16\pi m_X^2}{\mu}\sqrt{2\mu \delta_X},
\end{eqnarray}
where $\delta_X$ is the binding energy, and $\mu$ is the reduced mass of the $\bar{D}^0$ and $D^{\ast 0}$, i.e., $\mu=m_{\bar{D}^0}m_{D^{*0}}/(m_{\bar{D}^0}+m_{D^{*0}})$. The above equation is valid for the bound state ($\delta_X>0$). For the resonant case, the coupling can be evaluated as the residue of the $\bar{D}D^*$ scattering $T$ matrix~\cite{Gamermann:2009fv}. The coupling $g_X$ only affects the strength of the rescattering amplitude but will not change the line shape of the distribution curve. We therefore take a moderate value $g_X=3~\rm{GeV}$ which corresponds to the $\delta_X$ is at the order of magnitude of 100 keV, as did in Ref.~\cite{Sakai:2020ucu}. We should also mention that, although the $g_X$ does not affect the line shape behavior, it intrinsically depends on the nature of $X(3872)$.

The decay amplitude of $B^0\to \pi^- K^+ X(3872)$ via the $D_{s1}\bar{D}D^*$ triangle loop figured in Fig. \ref{triangleDiagram} reads
\begin{eqnarray}\label{amptT}
	&&\mathcal{M}=i\int \frac{d^4 q}{\left(2\pi\right)^4}\left(\frac{g_X}{2}  g_{D_{s1}D^*K}  \mathcal{C}_1  \right) \left(\vec{p}_\pi  \cdot\vec{\epsilon}^*(X) \right) \times \nonumber\\
	&&\frac{1}{\left(\left(P-q\right)^2 -m^2_{D_{s1}}\right) \left(q^2 -m^2_{\bar{D}}\right) \left(\left(p_X -q\right)^2-m^2_{D^{\ast }}\right)}. \nonumber \\
\end{eqnarray}
For the spin-1 state, the sum over polarization takes the form $\sum \epsilon_i \epsilon_j^*=\delta_{ij}$.
It should be mentioned that we adopt the non-relativistic amplitudes in Eqs.~(\ref{vertex1}), (\ref{Ds1DstarK}) and (\ref{XDDstar}), but we do not take the non-relativistic approximations for the denominators of the three propagators in the loop integral as shown in Eq.~(\ref{amptT}), since the formalism of these vertexes does not affect the line shape behaviour around the threshold we are interested in. The line shape of the distribution curve mainly depends the loop integral, which is numerically evaluated by employing the program package \textit{LoopTools}~\cite{Hahn:2000jm}.

The partial decay width of $B^0\to \pi^- K^+ X(3872)$ reads
\begin{eqnarray}\label{dGamma}
	\frac{d \Gamma_{B\to \pi K X}}{d M_{K X}}=\frac{p_K \tilde{p}_{\pi}}{\left(2\pi\right)^3 4m_B^2} \vert \mathcal{M}\vert ^2,
\end{eqnarray}
where
\begin{eqnarray}\label{phase}
	p_K&=& \frac{1}{2M_{KX}}\lambda^{1/2}\left(M^2_{KX}, m_K^2, m^2_X\right),\\
	\tilde{p}_{\pi}&=& \frac{1}{2m_B}\lambda^{1/2}\left(m_B^2, m_{\pi}^2,M_{KX}^2\right),
\end{eqnarray}
with $\lambda\left(x,y,z\right)=x^2+y^2+z^2 -2xy -2yz-2zx$.

The TS is a logarithmic singularity. To avoid the infinity of the loop integral in the physical region, one can replace the Feynman's $i\epsilon$ for the propagator by $im \Gamma$ with $\Gamma$ the total decay width, or equivalently replace the real mass $m$ by the complex mass $m-i\Gamma/2$, which will remove the TS from the physical boundary by a small distance~\cite{Aitchison:1964rwb,Denner:2014zga,Denner:2006ic}. The physical meaning of this complex mass prescription for avoiding the infinity is obvious. As long as the kinematic conditions for the TS being present on the physical boundary are fulfilled, it implies that the intermediate state (here is $D_{s1}(2536)$) is unstable, and it is necessary to take the finite width effect into account. Correspondingly, we replace the mass $m_{D_{s1}}$ in Eq.~(\ref{amptT}) by $m_{D_{s1}}-i\Gamma_{D_{s1}}/2$. The central values $m_{D_{s1}}=2535.11$ MeV and $\Gamma_{D_{s1}}=0.92$ MeV from Ref.~\cite{PDG:2022} are adopted in the numerical calculations.

\begin{figure}[htb]
	\centering
	\includegraphics[width=0.45\textwidth]{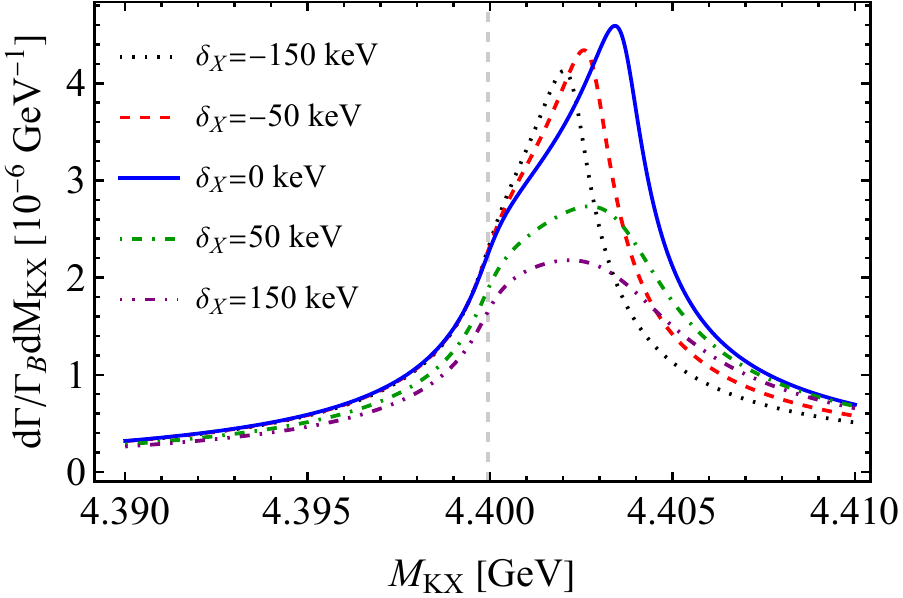}
	\caption{The $K^+ X(3872)$ invariant mass distributions around  $D_{s1}(2536)\bar{D}^0$ threshold (vertical dashed line) via the rescattering process in Fig.~\ref{triangleDiagram}. Different curves correspond to different $X(3872)$ masses. }\label{fig:0width}
\end{figure}

The invariant mass distributions of $K^+X(3872)$ via the triangle diagram in Fig.~\ref{triangleDiagram} are displayed in Fig.~\ref{fig:0width}. The $X(3872)$ mass is varied in a window $\delta_X \in [-150,\ 150]$ keV.
One can see that for different $m_X$ or $\delta_X$, the line shapes are also quite different. For every distribution curve in Fig.~\ref{fig:0width}, there is a cusp just at the $D_{s1}(2536)\bar{D}^0$ threshold. But these threshold cusps are smeared to some extent by the width effect of $D_{s1}(2536)$. The peak looks more clear and narrower for the negative $\delta_X$ compared with that for the positive $\delta_X$. Supposing the masses of intermediate states are real,  if $m_X$ is larger than or equal to the $D^{*0}\bar{D}^0$ threshold ($\delta_X\le 0$), the TS in $M_{KX}$ can be present on the physical boundary, and the corresponding TS peak in the distribution curve can be very sharp and the peak position is a little bit higher than the $D_{s1}(2536)\bar{D}^0$ threshold, but if $m_X$ is smaller than the $D^{*0}\bar{D}^0$ threshold ($\delta_X > 0$), the conditions of TS in $M_{KX}$ being present on the physical boundary can never be fulfilled, and one does not expect a sharp peak to appear in the distribution curve. Just because of this special character, the line shapes of $M_{KX}$ spectrum are very sensitive to the $m_X$. Especially, one can easily distinguish whether the $\delta_X$ is positive or negative by measuring the $K^+ X(3872)$ spectrum.

For the peaks shown in Fig.~\ref{fig:0width}, if we define a ``width'' at half maximum of the line shape, we can see the width can be as large as 3 to 5 MeV. Although this width is not a well defined quantity because of the asymmetric line shape, in terms of which we can still see an advantage for experiment: the larger width of the TS peak may reduce the requirement for energy resolution in measuring the line shape. The larger width is related to the larger TS window as shown in Eq.~(\ref{eq:3}).

\begin{figure}[tb]
	\centering
	\includegraphics[width=0.42\textwidth]{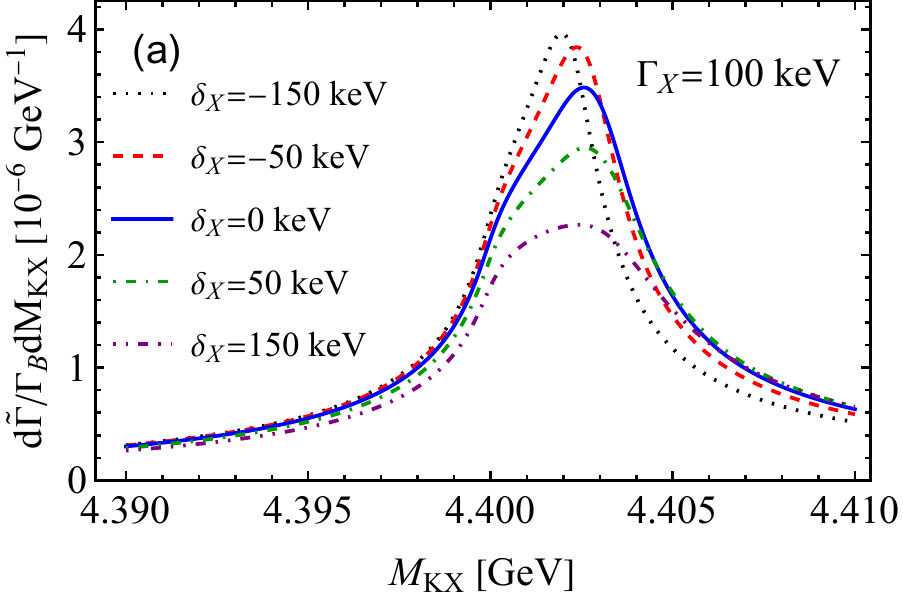}
	\includegraphics[width=0.42\textwidth]{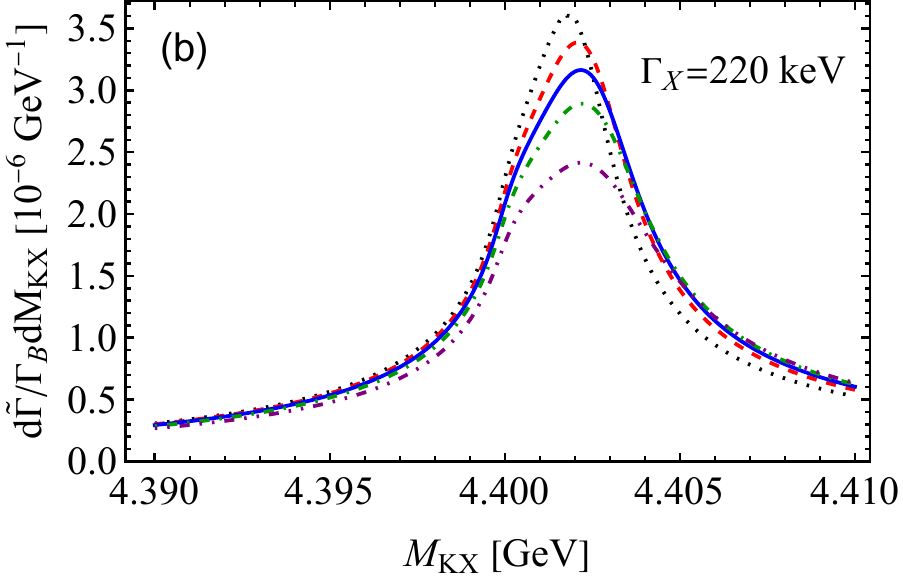} 
	\includegraphics[width=0.42\textwidth]{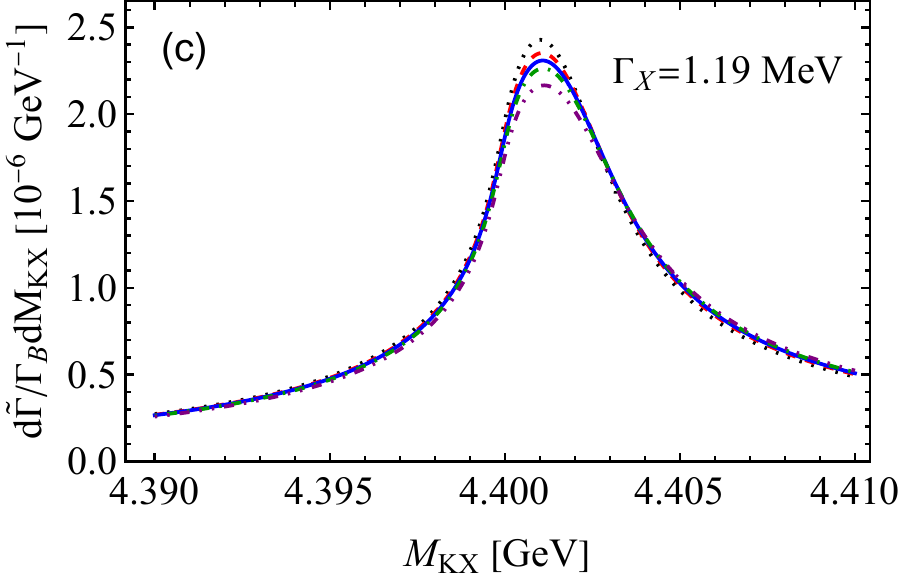}  
	\caption{The $X(3872)$ width dependent $K^+ X(3872)$ invariant mass distributions as defined in Eq.~(\ref{widthSpectral})  via the rescattering process in Fig.~\ref{triangleDiagram}. The width of $X(3872)$ is fixed to be (a) 100 keV, (b) 220 keV and (c) 1.19 MeV, respectively. The dotted, dashed, solid, dot-dashed and dot-dot-dashed curves corresponds to $\delta_X=$-150, -50, 0, 50 and 150 keV, respectively. }\label{fig:width}
\end{figure}

\subsection{Width impact of the $X(3872)$}

The $X(3872)$ is not a stable particle, and we need to take into account its width impact on the $KX(3872)$ line shape we are interested in. The Belle collaboration reports the branching fraction $\mathcal{B}(X(3872)\to D^0\bar{D}^{*0})$ is about $37\%$~\cite{Belle:2008fma}, and the $\mathcal{B}(X(3872)\to D^0\bar{D}^{0}\pi^0)$ is about $40\%$ with large uncertainties~\cite{Belle:2006olv}. The partial decay width $\Gamma(X(3872)\to D^0\bar{D}^{0}\pi^0)$ is expected to be about 40 keV in literatures~\cite{Dai:2019hrf,Guo:2014hqa,Fleming:2007rp}. Then the total width of $X(3872)$ can be estimated at around 100 keV. On the other hand, in Ref.~\cite{LHCb:2020xds}, the LHCb collaboration reports the Breit-Wigner (BW) width of $X(3872)$ is $\Gamma_{BW}=1.39\pm 0.24\pm 0.10$ MeV. But considering that the proximity of $X(3872)$ mass to the $ D^0\bar{D}^{*0}$ threshold may distort the line shape from the simple BW form, the LHCb also reports the full width at half maximum (FWHM) of the line shape is $\Gamma_{FWHM}=0.22^{+0.07+0.11}_{-0.06-0.13} $ MeV by using a Flatt{\'e}-inspired model~\cite{LHCb:2020xds}. One can see that the width value is subtle and highly depends on the fitting methods for the near threshold state $X(3872)$. The PDG 2022 gives the averaged BW width $\Gamma_{X} =1.19\pm 0.21\ \mbox{MeV}$~\cite{PDG:2022}. 
We take into account the $X(3872)$ width impact on the $KX$ line shape by introducing a new invariant mass distribution function
\begin{eqnarray}\label{widthSpectral}
	\frac{d \tilde{\Gamma}_{B\to \pi K X}}{d M_{K X}}&=&\int \limits^{(m_X+2\Gamma_X)^2}_{(m_X -2\Gamma_{X})^2}  dm^2 \rho_X(m^2)\frac{d \Gamma_{B\to \pi K X}}{d M_{KX}},
\end{eqnarray}
where the spectral function $\rho_X$ is defined to be
\begin{eqnarray}\label{spectral}
	\rho_X&=& \frac{1}{\mathcal{N}}\left(-\frac{1}{\pi}\right){\rm{Im}}\left[\frac{1}{m^2-m_X^2 + i m_X \Gamma_X}\right],
\end{eqnarray}
with
\begin{equation}
	\mathcal{N}= \int \limits^{(m_X+2\Gamma_X)^2}_{(m_X -2\Gamma_{X})^2} d m^2 \left(-\frac{1}{\pi}\right) {\rm{Im}} \left[\frac{1}{m^2-M_X^2 + i M_X \Gamma_X}\right].
\end{equation}
The same functions are adopted in Ref.~\cite{Sakai:2020ucu}. Another spectral function with Flatt{\'e} parametrization is adopted in Ref.~\cite{Sakai:2020crh}. The corresponding new distribution curves are displayed in Fig.~\ref{fig:width}. Compared with Fig.~\ref{fig:0width}, we can see that the strengths are weakened and the curves are smoothed to some extent. When $\Gamma_X$ is set to be 100 keV and 220 keV, one can still see relatively larger discrepancy between the curves corresponding to negative and positive $\delta_X$. But when $\Gamma_X$ is set to be 1.19 MeV, the discrepancy between different curves is tiny and those curves nearly overlap with each other, which implies the sensitiveness of the line shape on $\delta_X$ is reduced when $\Gamma_X$ is larger. From this point of view, we comment that if the width of $X(3872)$ is as large as 1 MeV, the TS mechanism of measuring its mass may be ruined.

If we use the TS mechanism to measure the discrepancy between $m_X$ and $D^{*0}\bar{D}^0$ threshold, the $X(3872)$ needs to be reconstructed in decay modes other than the $D^0\bar{D}^0\pi^0$, which can be $J/\psi\pi^+\pi^-$, $J/\psi\pi^+\pi^- \pi^0$ and so on. Otherwise one has to take into account the interference term between the rescattering triangle diagram and the cascade decay process $B^0\to\pi^- D_{s1}(2536)^+ \bar{D}^0 \to \pi^- K^+ D^0 \bar{D}^0\pi^0 $. This interference is subtle and the tree level cascade decay process cannot be treated as a smooth background near the TS regions because of the so-called Schmid theorem~\cite{Schmid:1967ojm,Aitchison:1969tq,Goebel:1982yb,Anisovich:1995ab,Debastiani:2018xoi}.  This point has ever been pointed out in Ref.~\cite{Guo:2019qcn}.


\section{Summary}
In summary, we investigate the $B^0\to \pi^- K^+ X(3872)$ decay via a  triangle rescattering diagram, where the $B^0$$ \to$$ \pi^-$$ D_{s1}(2536)^+ \bar{D}^0$ decay followed by $D_{s1}(2536)^+$ decaying into $ D^{\ast 0}K^+$ and $D^{\ast 0}\bar{D}^0$ fusing into $X(3872)$. The TS of rescattering amplitude can be generated from the $D_{s1}(2536)\bar{D} D^*$ loop, and the line shape of $K^+X(3872)$ distribution curve is very sensitive to the $X(3872)$ mass. By means of this characteristic, we can determine whether the $X(3872)$ mass is below or above $D^{\ast 0}\bar{D}^0$ threshold, which is crucial in understanding the nature of $X(3872)$. The narrowness of $D_{s1}(2536)$ in this $D_{s1}\bar{D} D^*$ loop is one of the key reasons why the TS mechanism of measuring the $X(3872)$ mass may work. The relatively larger TS kinematic window may also reduce the experimental requirement for energy resolution. This indirect method of measuring the $X(3872)$ mass in $B^0\to \pi^- K^+ X(3872)$ decay via the $D_{s1}\bar{D} D^*$ loop may be feasible in the LHCb and updated Belle II experiments.

We also take into account the $X(3872)$ width impact on the $KX$ line shape by introducing a distribution function convoluted with the $X(3872)$ spectral function. It is shown that for the $X(3872)$ width to be at the order of 100 keV, the influence of the width is small. But if the $X(3872)$ width is as large as 1 MeV, the method of using the TS mechanism to precisely measure its mass would be ruined.

\begin{acknowledgments}
This work is supported in part by the National Natural Science Foundation of China (NSFC) under Grants No.~11975165, No. 11835015, No. 12047503, No. 12125507,  the Chinese Academy of Sciences under Grant No. XDPB15 and the China Postdoctoral Science Foundation under Grant No. 2022M713229.
\end{acknowledgments}


\begin{thebibliography}{99}

\bibitem{Guo:2019twa}
F.~K.~Guo, X.~H.~Liu and S.~Sakai,
Prog. Part. Nucl. Phys. \textbf{112}, 103757 (2020)
doi:10.1016/j.ppnp.2020.103757
[arXiv:1912.07030 [hep-ph]].


\bibitem{PDG:2022}
R.~L. Workman {\it et al.} (Particle Data Group), Prog. Theor. Exp. Phys. 2022, 083C01 (2022)



\bibitem{Guo:2017jvc}
F.~K.~Guo, C.~Hanhart, U.~G.~Mei\ss{}ner, Q.~Wang, Q.~Zhao and B.~S.~Zou,
Rev. Mod. Phys. \textbf{90}, no.1, 015004 (2018)
[erratum: Rev. Mod. Phys. \textbf{94}, no.2, 029901 (2022)]
doi:10.1103/RevModPhys.90.015004
[arXiv:1705.00141 [hep-ph]].

\bibitem{Kalashnikova:2018vkv}
Y.~S.~Kalashnikova and A.~V.~Nefediev,
Phys. Usp. \textbf{62}, no.6, 568-595 (2019)
doi:10.3367/UFNe.2018.08.038411
[arXiv:1811.01324 [hep-ph]].

\bibitem{Brambilla:2019esw}
N.~Brambilla, S.~Eidelman, C.~Hanhart, A.~Nefediev, C.~P.~Shen, C.~E.~Thomas, A.~Vairo and C.~Z.~Yuan,
Phys. Rept. \textbf{873}, 1-154 (2020)
doi:10.1016/j.physrep.2020.05.001
[arXiv:1907.07583 [hep-ex]].

\bibitem{Chen:2016qju}
H.~X.~Chen, W.~Chen, X.~Liu and S.~L.~Zhu,
Phys. Rept. \textbf{639}, 1-121 (2016)
doi:10.1016/j.physrep.2016.05.004
[arXiv:1601.02092 [hep-ph]].

\bibitem{Esposito:2016noz}
A.~Esposito, A.~Pilloni and A.~D.~Polosa,
Phys. Rept. \textbf{668}, 1-97 (2017)
doi:10.1016/j.physrep.2016.11.002
[arXiv:1611.07920 [hep-ph]].

\bibitem{Olsen:2017bmm}
S.~L.~Olsen, T.~Skwarnicki and D.~Zieminska,
Rev. Mod. Phys. \textbf{90}, no.1, 015003 (2018)
doi:10.1103/RevModPhys.90.015003
[arXiv:1708.04012 [hep-ph]].

\bibitem{Lebed:2016hpi}
R.~F.~Lebed, R.~E.~Mitchell and E.~S.~Swanson,
Prog. Part. Nucl. Phys. \textbf{93}, 143-194 (2017)
doi:10.1016/j.ppnp.2016.11.003
[arXiv:1610.04528 [hep-ph]].

\bibitem{Ali:2017jda}
A.~Ali, J.~S.~Lange and S.~Stone,
Prog. Part. Nucl. Phys. \textbf{97}, 123-198 (2017)
doi:10.1016/j.ppnp.2017.08.003
[arXiv:1706.00610 [hep-ph]].

\bibitem{Liu:2019zoy}
Y.~R.~Liu, H.~X.~Chen, W.~Chen, X.~Liu and S.~L.~Zhu,
Prog. Part. Nucl. Phys. \textbf{107}, 237-320 (2019)
doi:10.1016/j.ppnp.2019.04.003
[arXiv:1903.11976 [hep-ph]].

\bibitem{Abe:2003hq}
K.~Abe \textit{et al.} [Belle],
[arXiv:hep-ex/0308029 [hep-ex]].

\bibitem{ATLAS:2016kwu}
M.~Aaboud \textit{et al.} [ATLAS],
JHEP \textbf{01}, 117 (2017)
doi:10.1007/JHEP01(2017)117
[arXiv:1610.09303 [hep-ex]].

\bibitem{CMS:2013fpt}
S.~Chatrchyan \textit{et al.} [CMS],
JHEP \textbf{04}, 154 (2013)
doi:10.1007/JHEP04(2013)154
[arXiv:1302.3968 [hep-ex]].

\bibitem{LHCb:2011zzp}
R.~Aaij \textit{et al.} [LHCb],
Eur. Phys. J. C \textbf{72}, 1972 (2012)
doi:10.1140/epjc/s10052-012-1972-7
[arXiv:1112.5310 [hep-ex]].

\bibitem{D0:2004zmu}
V.~M.~Abazov \textit{et al.} [D0],
Phys. Rev. Lett. \textbf{93}, 162002 (2004)
doi:10.1103/PhysRevLett.93.162002
[arXiv:hep-ex/0405004 [hep-ex]].

\bibitem{Guo:2019qcn}
F.~K.~Guo,
Phys. Rev. Lett. \textbf{122}, no.20, 202002 (2019)
doi:10.1103/PhysRevLett.122.202002
[arXiv:1902.11221 [hep-ph]].

\bibitem{Sakai:2020crh}
S.~Sakai, H.~J.~Jing and F.~K.~Guo,
Phys. Rev. D \textbf{102}, no.11, 114041 (2020)
doi:10.1103/PhysRevD.102.114041
[arXiv:2008.10829 [hep-ph]].

\bibitem{Braaten:2019gfj}
E.~Braaten, L.~P.~He and K.~Ingles,
Phys. Rev. D \textbf{100}, no.3, 031501 (2019)
doi:10.1103/PhysRevD.100.031501
[arXiv:1904.12915 [hep-ph]].

\bibitem{Braaten:2019gwc}
E.~Braaten, L.~P.~He and K.~Ingles,
Phys. Rev. D \textbf{101}, no.1, 014021 (2020)
doi:10.1103/PhysRevD.101.014021
[arXiv:1909.03901 [hep-ph]].

\bibitem{Braaten:2019yua}
E.~Braaten, L.~P.~He and K.~Ingles,
Phys. Rev. D \textbf{100}, no.7, 074028 (2019)
doi:10.1103/PhysRevD.100.074028
[arXiv:1902.03259 [hep-ph]].

\bibitem{Sakai:2020ucu}
S.~Sakai, E.~Oset and F.~K.~Guo,
Phys. Rev. D \textbf{101}, no.5, 054030 (2020)
doi:10.1103/PhysRevD.101.054030
[arXiv:2002.03160 [hep-ph]].

\bibitem{Braaten:2022elw}
E.~Braaten, L.~P.~He, K.~Ingles and J.~Jiang,
[arXiv:2202.03900 [hep-ph]].


\bibitem{ParticleDataGroup:2020ssz}
P.~A.~Zyla \textit{et al.} [Particle Data Group],
PTEP \textbf{2020}, no.8, 083C01 (2020)
doi:10.1093/ptep/ptaa104

\bibitem{Belle:2015qeg}
A.~Bala \textit{et al.} [Belle],
Phys. Rev. D \textbf{91}, no.5, 051101 (2015)
doi:10.1103/PhysRevD.91.051101
[arXiv:1501.06867 [hep-ex]].

\bibitem{Coleman:1965xm}
S.~Coleman and R.~E.~Norton,
Nuovo Cim. \textbf{38}, 438-442 (1965)
doi:10.1007/BF02750472

\bibitem{Landau:1959fi}
L.~D.~Landau,
Nucl. Phys. \textbf{13}, no.1, 181-192 (1959)
doi:10.1016/B978-0-08-010586-4.50103-6

\bibitem{Liu:2015taa}
X.~H.~Liu, M.~Oka and Q.~Zhao,
Phys. Lett. B \textbf{753}, 297-302 (2016)
doi:10.1016/j.physletb.2015.12.027
[arXiv:1507.01674 [hep-ph]].

\bibitem{Bronzan:1963mby}
J.~B.~Bronzan and C.~Kacser,
Phys. Rev. \textbf{132}, no.6, 2703 (1963)
doi:10.1103/PhysRev.132.2703

\bibitem{Aitchison:1964zz}
I.~J.~R.~Aitchison,
Phys. Rev. \textbf{133}, B1257-B1266 (1964)
doi:10.1103/PhysRev.133.B1257

\bibitem{Weinberg:1965zz}
S.~Weinberg,
Phys. Rev. \textbf{137}, B672-B678 (1965)
doi:10.1103/PhysRev.137.B672

\bibitem{Baru:2003qq}
V.~Baru, J.~Haidenbauer, C.~Hanhart, Y.~Kalashnikova and A.~E.~Kudryavtsev,
Phys. Lett. B \textbf{586}, 53-61 (2004)
doi:10.1016/j.physletb.2004.01.088
[arXiv:hep-ph/0308129 [hep-ph]].

\bibitem{Gamermann:2009uq}
D.~Gamermann, J.~Nieves, E.~Oset and E.~Ruiz Arriola,
Phys. Rev. D \textbf{81}, 014029 (2010)
doi:10.1103/PhysRevD.81.014029
[arXiv:0911.4407 [hep-ph]].

\bibitem{Gamermann:2009fv}
D.~Gamermann and E.~Oset,
Phys. Rev. D \textbf{80}, 014003 (2009)
doi:10.1103/PhysRevD.80.014003
[arXiv:0905.0402 [hep-ph]].

\bibitem{Hahn:2000jm}
T.~Hahn,
Nucl. Phys. B Proc. Suppl. \textbf{89}, 231-236 (2000)
doi:10.1016/S0920-5632(00)00848-3
[arXiv:hep-ph/0005029 [hep-ph]].

\bibitem{Aitchison:1964rwb}
I.~J.~R.~Aitchison and C.~Kacser,
Phys. Rev. \textbf{133}, no.5B, B1239-B1257 (1964)
doi:10.1103/physrev.133.b1239

\bibitem{Denner:2014zga}
A.~Denner and J.~N.~Lang,
Eur. Phys. J. C \textbf{75}, no.8, 377 (2015)
doi:10.1140/epjc/s10052-015-3579-2
[arXiv:1406.6280 [hep-ph]].

\bibitem{Denner:2006ic}
A.~Denner and S.~Dittmaier,
Nucl. Phys. B Proc. Suppl. \textbf{160}, 22-26 (2006)
doi:10.1016/j.nuclphysbps.2006.09.025
[arXiv:hep-ph/0605312 [hep-ph]].

\bibitem{Belle:2008fma}
T.~Aushev \textit{et al.} [Belle],
Phys. Rev. D \textbf{81}, 031103 (2010)
doi:10.1103/PhysRevD.81.031103
[arXiv:0810.0358 [hep-ex]].

\bibitem{Belle:2006olv}
G.~Gokhroo \textit{et al.} [Belle],
Phys. Rev. Lett. \textbf{97}, 162002 (2006)
doi:10.1103/PhysRevLett.97.162002
[arXiv:hep-ex/0606055 [hep-ex]].

\bibitem{Dai:2019hrf}
L.~Dai, F.~K.~Guo and T.~Mehen,
Phys. Rev. D \textbf{101}, no.5, 054024 (2020)
doi:10.1103/PhysRevD.101.054024
[arXiv:1912.04317 [hep-ph]].

\bibitem{Guo:2014hqa}
F.~K.~Guo, C.~Hidalgo-Duque, J.~Nieves, A.~Ozpineci and M.~P.~Valderrama,
Eur. Phys. J. C \textbf{74}, no.5, 2885 (2014)
doi:10.1140/epjc/s10052-014-2885-4
[arXiv:1404.1776 [hep-ph]].

\bibitem{Fleming:2007rp}
S.~Fleming, M.~Kusunoki, T.~Mehen and U.~van Kolck,
Phys. Rev. D \textbf{76}, 034006 (2007)
doi:10.1103/PhysRevD.76.034006
[arXiv:hep-ph/0703168 [hep-ph]].

\bibitem{LHCb:2020xds}
R.~Aaij \textit{et al.} [LHCb],
Phys. Rev. D \textbf{102}, no.9, 092005 (2020)
doi:10.1103/PhysRevD.102.092005
[arXiv:2005.13419 [hep-ex]].

\bibitem{Schmid:1967ojm}
C.~Schmid,
Phys. Rev. \textbf{154}, no.5, 1363 (1967)
doi:10.1103/PhysRev.154.1363

\bibitem{Aitchison:1969tq}
I.~J.~R.~Aitchison and C.~Kacser,
Phys. Rev. \textbf{173}, 1700-1708 (1968)
doi:10.1103/PhysRev.173.1700

\bibitem{Goebel:1982yb}
C.~J.~Goebel, S.~F.~Tuan and W.~A.~Simmons,
Phys. Rev. D \textbf{27}, 1069 (1983)
doi:10.1103/PhysRevD.27.1069

\bibitem{Anisovich:1995ab}
A.~V.~Anisovich and V.~V.~Anisovich,
Phys. Lett. B \textbf{345}, 321-324 (1995)
doi:10.1016/0370-2693(94)01671-X

\bibitem{Debastiani:2018xoi}
V.~R.~Debastiani, S.~Sakai and E.~Oset,
Eur. Phys. J. C \textbf{79}, no.1, 69 (2019)
doi:10.1140/epjc/s10052-019-6558-1
[arXiv:1809.06890 [hep-ph]].
	
	
\end{thebibliography}
\end{document}